\documentclass[a4paper,11pt]{article}
\usepackage{pos}
\usepackage{color}

\newcommand{\nob}[1]{\textcolor{black}{ #1}} 

\title{Probing \nob{P}article \nob{A}cceleration through \nob{G}amma-ray Solar \nob{F}lare \nob{O}bservations}
 \ShortTitle{Gamma-ray Solar flares}

\author*[a]{Melissa Pesce-Rollins}
\author[b]{Nicola Omodei}
\author[b]{Vahe' Petrosian}
\author[c]{Francesco Longo}

\affiliation[a]{Istituto Nazionale di Fisica Nucleare, Sezione di Pisa, \\
  I-56127 Pisa, Italy}

\affiliation[b]{W. W. Hansen Experimental Physics Laboratory, Kavli Institute for Particle Astrophysics and Cosmology, Department of Physics and SLAC National Accelerator
Laboratory, Stanford University,\\
 Stanford, CA 94305, USA}

\affiliation[c]{Department of Physics, University of Trieste and INFN, sezione di Trieste, via Valerio 2,\\
  I-34127 Trieste, Italy}

\emailAdd{melissa.pesce.rollins@pi.infn.it}

\abstract{High-energy solar flares have shown to have at least two distinct phases: prompt-impulsive and delayed-gradual. Identifying the mechanism responsible for accelerating the electrons and ions and the site at which it occurs during these two phases is one of the outstanding questions in solar physics. Many advances have been made over the past decade thanks to new observational data and refined simulations that together help to shed light on this topic. For example, the detection by Fermi Large Area Telescope (LAT) of GeV emission from solar flares originating from behind the visible solar limb and >100 MeV emission lasting for more than 20 hours have suggested the need for a spatially extended source of acceleration during the delayed emission phase. In this work we will review some of the major results from \emph{Fermi}-LAT observations of the 24$^{th}$ solar cycle and how this new observational channel combined with observations from across the electromagnetic spectrum can  provide a unique opportunity to diagnose the mechanisms of high-energy emission and particle acceleration in solar flares. }

\FullConference{37$^{\rm{th}}$ International Cosmic Ray Conference (ICRC 2021)\\
		July 12th -- 23rd, 2021\\
		Online -- Berlin, Germany}

\begin{document}
\maketitle

\section{Introduction}
\label{sec:intro}
Solar flares are explosive phenomena that emit electromagnetic radiation over an extremely wide range covering from radio to gamma rays. It is generally accepted that the magnetic energy stored in the solar corona and released through reconnection during a flare is capable of accelerating electrons and ions to relativistic energies.  Much is known of the electron acceleration thanks to the observations made in Hard X-rays (10 keV -- 1 MeV; HXRs;~\citep[see, e.g.~][]{Vilmer1987,1988SoPh..118...49D,LIN20031001}\nob{)} and microwaves \citep[see, e.g.~][]{1998A&A...334.1099T}. In general, the gamma-ray emission light curve is similar to that of the HXRs, lasting for 10-100 s. This is referred to as the prompt-impulsive phase of the flare. The detection by the {\it Energetic Gamma Ray Experiment Telescope} (EGRET) on-board CGRO~\citep{Kanbach:88,Esposito:99} of gamma rays above 100 MeV for more than an hour after the impulsive phases of \nob{three} flares \citep{2000SSRv...93..581R} illustrated how another extended phase was present for the gamma-ray component. In particular, the 1991 June 11 flare was remarkable because the gamma-ray emission ($>$50 MeV) lasted for 8 hours after the impulsive phase of the flare~\citep{1993A&AS...97..349K} and thus the term \emph{Long Duration Gamma-ray Flare} was born. Unfortunately limited observations of such extended phase emission have left several open questions to be answered. With the launch of the Large Area Telescope (LAT)~\cite{LATPaper} on-board the \emph{Fermi} Gamma-ray Space Telescope the number of high-energy gamma-ray flares observed has substantially increased\nob{,} allowing for a better understanding of the acceleration mechanisms during both the prompt and delayed phases of solar flares.

In this work a brief review of the observational signatures of particle acceleration during solar flares and the information on the underlying particle population that can be learned from studying the gamma-ray emission from these explosive phenomena is presented.

\section{Accelerated particle signatures: Flare site}
\label{sec:acc}
\begin{figure}
  \includegraphics[width=.45\textwidth]{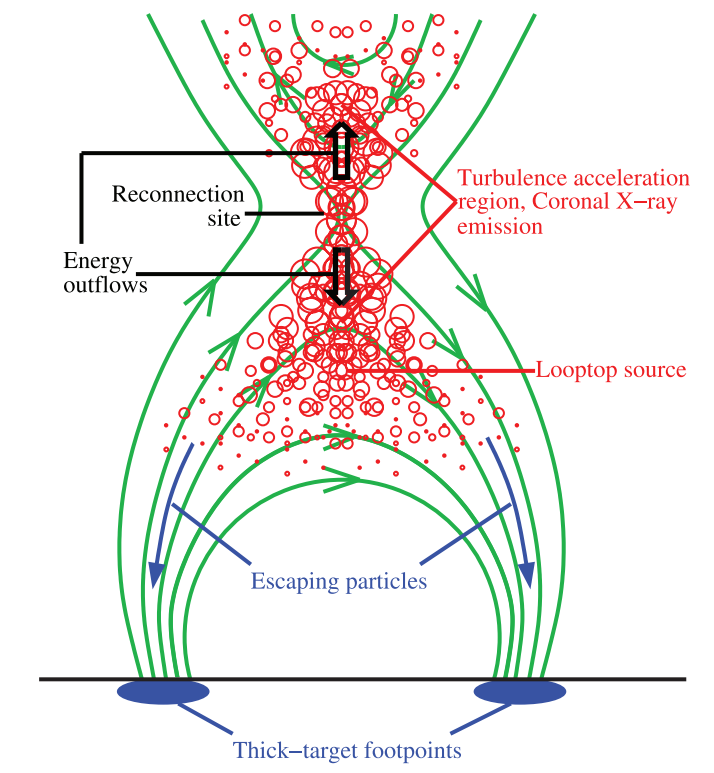}
  \includegraphics[width=.55\textwidth]{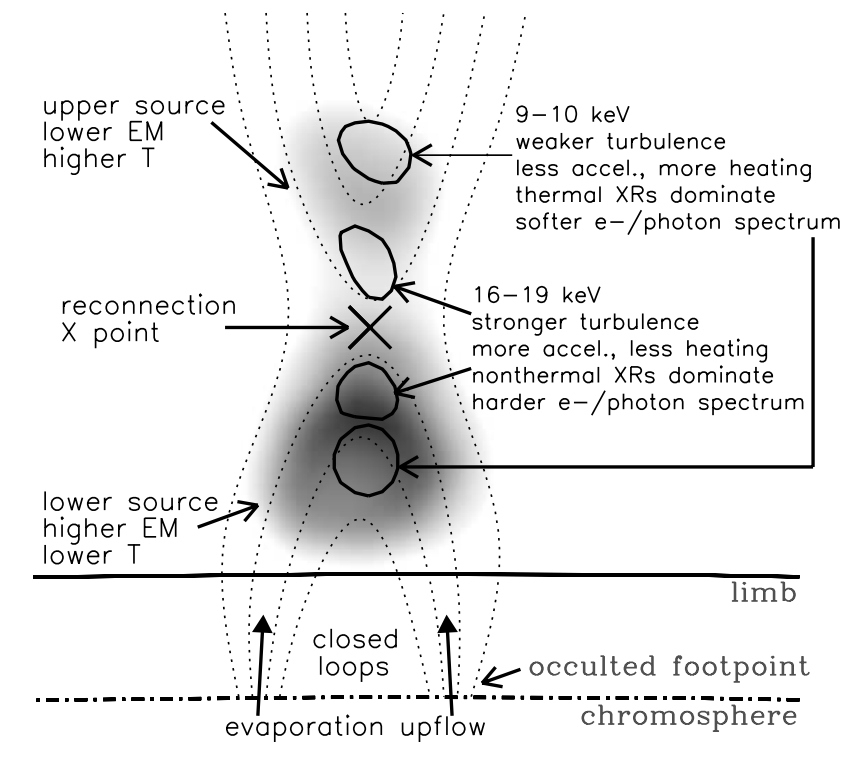}
\caption{Left panel: Sketch of the stochastic acceleration model (see text for references) proposed for solar flares. The green lines represent magnetic field lines in a possible configuration; the red circles represent turbulence or plasma waves that are generated during magnetic reconnection. Right panel: Schematic of the physical scenario with RHESSI observations of simultaneous 9-10 (thin) and 16-19 keV (thick) contours superimposed. Credits:~\cite{Liu_2008}}
\label{fig1}
\end{figure}

The main source of energy release and consequently for the particle acceleration is believed to be magnetic reconnection (see the reviews by~\cite{2011SSRv..159...19F, holman2016,2017LRSP...14....2B}). There are several models proposed for explaining the particle acceleration that occurs during solar flares (\citep[see, e.g.~][]{1994ApJ...435..469B,1993SoPh..146..127L,1996ApJ...461..445M}) and a review of the various models proposed can be found here~\cite{2011SSRv..159..357Z}. In the left-hand panel of Figure~\ref{fig1} is a cartoon representation of one the proposed models, the stochastic acceleration model (\citep[see, e.g.~][]{1992ApJ...398..350H,1995ApJ...446..699P,2004ApJ...610..550P,2012SSRv..173..535P}). In this acceleration model, once reconnection occurs, turbulence resulting from the energy outflows accelerate particles and soft and hard X-ray emission is expected both close to the reconnection region as well as at the footpoints of the magnetic field lines. This second source of emission is caused by the accelerated particles escaping the turbulent regions and depositing their energy in the denser regions of the Sun. These footpoints (also known as \emph{thick target sources}) are typically located in a compact regions around the active region (AR) from which the activity erupted and are more brighter than the looptop sources.

Imaging and spectral analyses in soft and hard X-rays provides supporting evidence for magnetic reconnection and acceleration in the corona~\citep[for a review, see, e.g.,][]{2011SSRv..159..107H}, in particular the double \nob{c}oronal sources reported by (e.g.,~\cite{Sui_2003,veronig_2006,Liu_2008,2013NatPh...9..489S}). In context of, for example, the stochastic acceleration model, two energy dependent sources should be visible tracing the level of turbulence generated by the outflows from the reconnection site (as shown on the left panel of Figure~\ref{fig1}). In the right-hand panel of Figure~\ref{fig1} are the RHESSI~\cite{2002SoPh..210....3L} observations of X-ray sources from the flare of 2002 April 30~\cite{Liu_2008} superimposed on the schematic of the stochastic model providing supporting evidence of magnetic reconnection occurring during a solar flare.

Solar flares are also sources of gamma-ray emission ranging from a few MeV to GeV energies. The emission is produced by interactions of high-energy particles with the ambient plasma through three main mechanisms: Bremsstrahlung (10~keV -- 1 GeV), nuclear de-excitation ($\approx$0.5 -- 8 MeV) and pion decay ($>$10 MeV) (see \cite{1998astro.ph.10089R,2000eaa..bookE2292R} for reviews of gamma ray production in solar flares). By studying the characteristics of the gamma-ray emission it is possible to obtain clues on the properties of the acceleration mechanisms and information of the ambient plasma.

\begin{figure}
  \includegraphics[width=.5\textwidth]{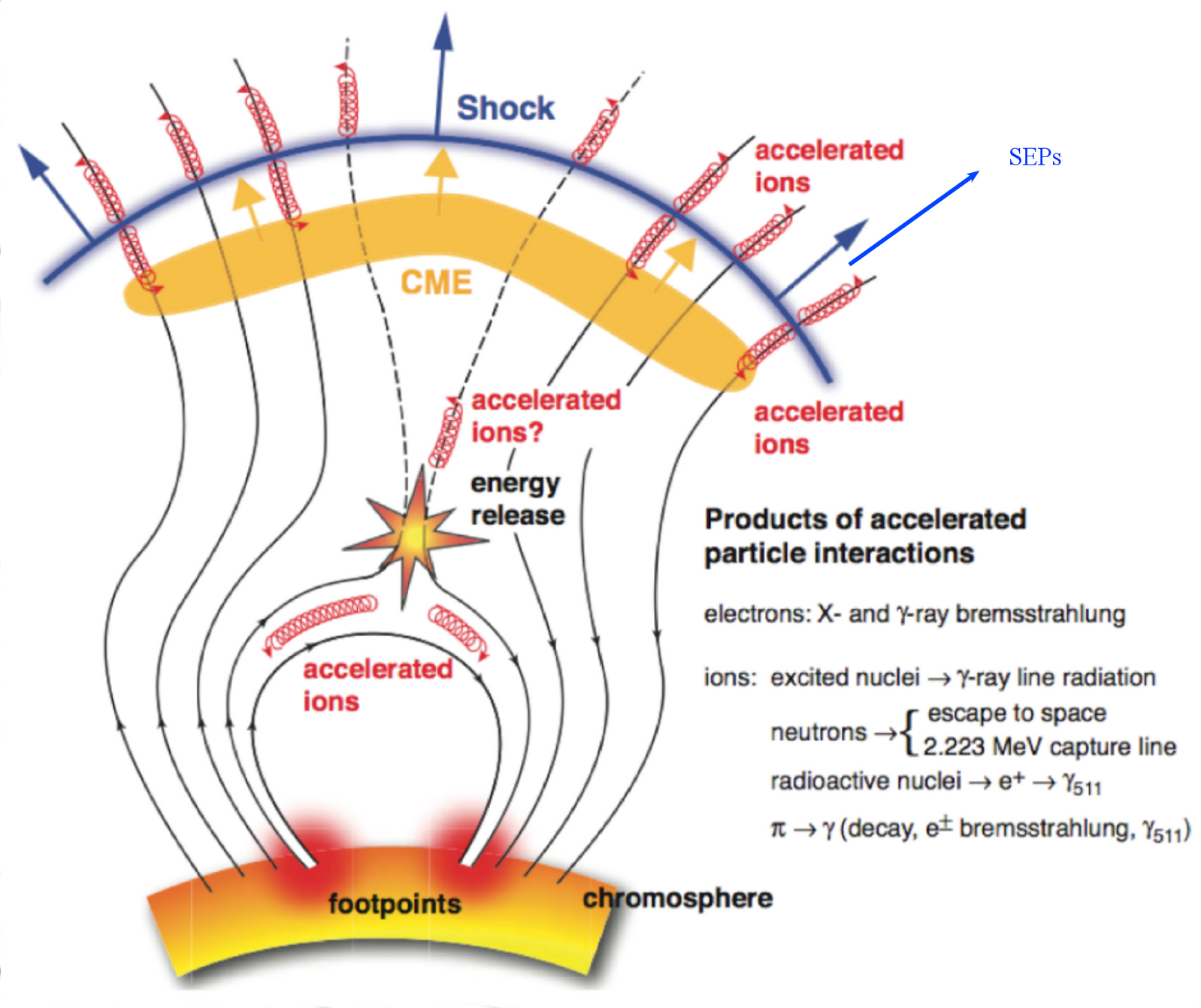}
  \includegraphics[width=.5\textwidth]{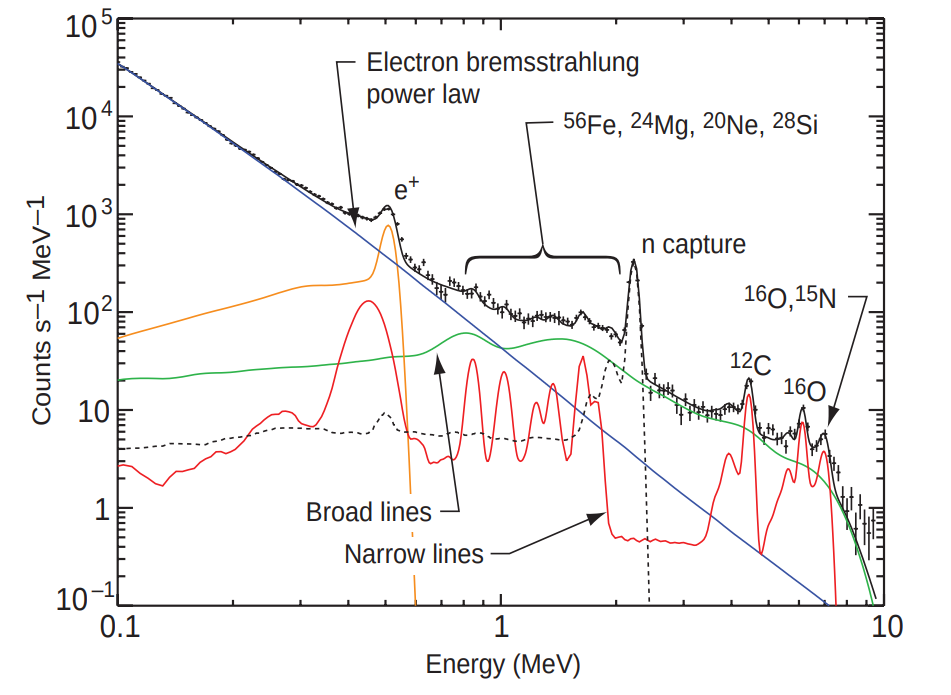}
  \caption{Left-hand panel: Cartoon illustrating the acceleration processes and the products of the accelerated particles occurring during solar flares. Right-hand panel: The spectral energy distribution of the gamma-ray flare of 1991 June 4 observed by CGRO-OSSE~\cite{1994AIPC..294...15M}. The deexcitation lines produced by the accelerated ions interaction with the ambient Solar material are annotated in the figure. The neutron capture line is also very evident.}
\label{fig2}
\end{figure}

The cartoon on the left panel of Figure~\ref{fig2}, demonstrates a possible magnetic connection between the flare and CME shock sites of acceleration. The particles traveling along the closed field lines precipitate down to the chromosphere to produce the footpoint emission (including the products of the accelerated ions) and how the particle traveling along the open field lines can escape from the flare site as Solar Energetic Particles (SEP).
The accelerated ions precipitate down to the chromosphere and interact with the
ambient Solar material exciting the nuclei (carbon, for example). These excited nuclei
promptly deexcite producing photons with the energy equal to the difference between the two excited states. Given the large number of ions precipitating during flares, gamma-ray lines become visible in the observed spectrum typically between 1 to 8 MeV. Another strong line in the spectrum of solar flares is the neutron capture line, this occurs because ion interactions also produce neutrons that can be captured on ambient hydrogen to produce deuterium. The binding energy of deuterium appears as a strong gamma-ray line at 2.223 MeV. The gamma-ray spectrum above $\sim$30 MeV is typically a continuum and is produced by neutral pions that decay directly into two 67.5 MeV gamma rays.

Thus observing the gamma-ray spectrum from $\sim$1 MeV up to GeV energies provides information on the population of ions between roughly 1 -- 300 MeV accelerated during the magnetic reconnection process~\cite{chup09}.  The right-hand panel of Figure~\ref{fig2} the spectral energy distribution of the gamma-ray flare of 1991 June 4 observed by CGRO-OSSE~\cite{1994AIPC..294...15M,murp97} is shown, where the line emission due to excited nuclei and neutrons are visible.


\section{Accelerated particle signatures: Far from the flare site}
\label{sec:acc_far}
The signatures for particle acceleration are observable not only from the compact region around the AR but also from regions far from the AR. Many intense flares are accompanied by Coronal Mass Ejections (CME) which are significant releases of plasma and magnetic field from the solar corona. These phenomena form shock waves that travel very quickly away from the Sun into interplanetary space. For decades now, there have been observations of SEP observed at Earth. These are particles originating from the Sun and appear to fall into to broad categories: impulsive and gradual. The acceleration mechanisms driving these two categories have been found to be associated with the flare acceleration and the CME-driven shock, respectively~\citep[see, e.g.~][]{2021LNP...978.....R,2012AAS...22011201R}. Observations of SEP from Earth have shown that ions are being accelerated up to GeV energies~\cite{2012SSRv..173..247L,tylk06}.
\begin{figure}
  \includegraphics[width=\textwidth]{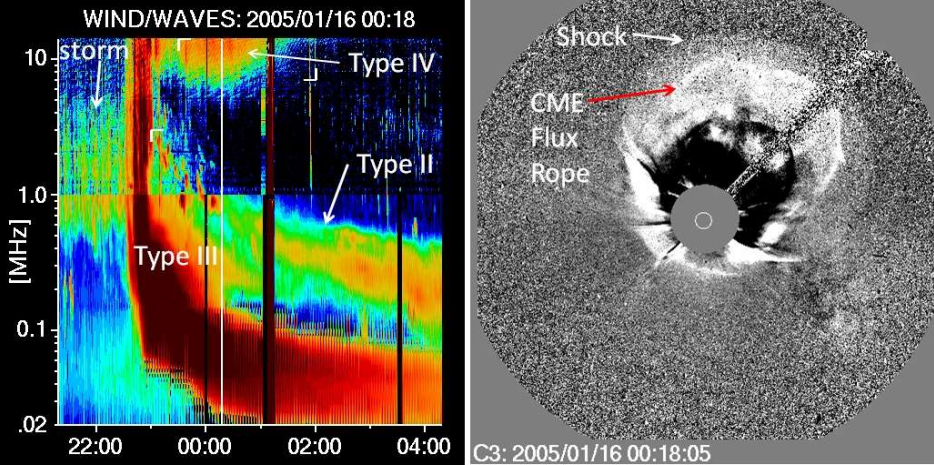}
\caption{Wind/WAVES dynamic spectrum (left) and the associated white-light CME observed
by SOHO/LASCO (right). The different radio bursts are labeled on the dynamic spectrum indicating the varying topologies. In the white-light image of the CME, the presence of the shock front is annotated. Credits~\cite{gopalswamy2019catalog}.}
\label{fig3}
\end{figure}


Other signatures of particle acceleration that have been found to be associated with CMEs are radio bursts~\cite{2019SunGe..14..111G}. The emission from these bursts, due to plasma emission, has been found to be a function of frequency and time and have been described in terms of five main types of bursts named type I through V~\cite{1963ARA&A...1..291W}.  The measured frequencies from these radio bursts are found to drift from higher to lower values with time. This is because the frequencies of these bursts depend on the electron density and magnetic field strength so as the shock propagates away from the Sun the electron density as well as the magnetic field decrease~\cite{bastian1998}. The frequency drift rate can be used to estimate the speed of the shock wave and has been found to match that of the associated CME speed determined from other methods~\citep[see, e.g.~][and references therein]{gopal2005}. In the left-hand panel of Figure~\ref{fig3} the dynamic spectrum of a radio burst measured by Wind/WAVES is shown, where three different types of radio bursts are \nob{detected}. The right-hand panel shows the associated white-light CME observed by the SOHO/LASCO detector showing the presence of the shock responsible for the radio emission. 


\section{Connection between X-rays and \nob{gamma} rays}
\label{xraytogamma}
\begin{figure}
  \includegraphics[width=.5\textwidth]{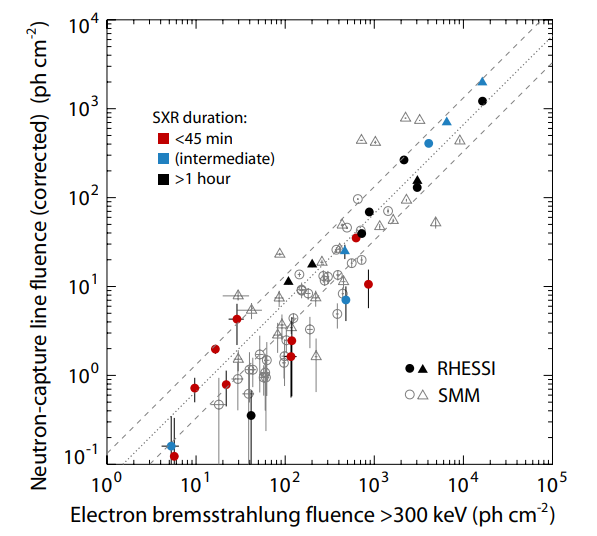}
  \includegraphics[width=.48\textwidth]{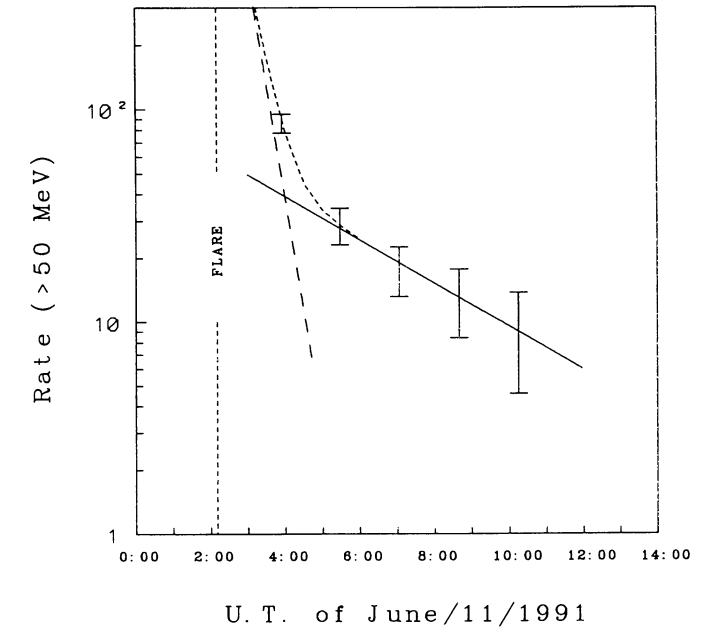}
\caption{Left panel: Fluence in the neutron-capture line versus the $>$0.3 MeV bremsstrahlung fluence for flares with heliocentric angles $<$80$^{\circ}$. The dotted line indicates the best-fit line in linear space that passes through the origin. See \cite{shih09} for further details. Right panel: Time profile for the $>$50 MeV emission from the flare of 1991 June 11 observed with with EGRET on COMPTON-GRO~\cite{1993A&AS...97..349K}. The time profile is fit with a two component model (indicated dashed and solid lines) with time constants of 25 and 255 minutes for the first and second component, respectively.}
\label{fig4}
\end{figure}

As discussed in Section~\ref{sec:acc}, there are imaging results supporting the stochastic acceleration scenario during the impulsive phase of solar flares. However, these results were referring to the acceleration of electrons but what can be said of the acceleration of ions during solar flares? Comparisons between the neutron-capture line fluence versus the $>$0.3 MeV bremsstrahlung fluence of solar flares observed with RHESSI and Gamma-Ray Spectrometer on SMM~\cite{1980SoPh...65...15F} indicate that the two fluences are tightly correlated (Pearson's correlation coefficient of r$^2 = 0.97$) over a range of $>$3 orders of magnitude~\cite{shih09} (also shown in the left panel of Figure~\ref{fig4}). This correlation over such a large range of fluence strongly implies a common acceleration mechanism between the relativistic electrons and the $>$30 MeV protons. Similar results have been reported between the $>$0.3~MeV electron bremsstrahlung emission and the 4-8 MeV nuclear line emission~\cite{1988SoPh..118...95V,murphy93}. A correlation of 4-8 MeV nuclear excess emission with $>$50 keV continuum emission has also been found~\cite{1994ApJ...426..767C}.
In the study reported in \cite{shih09}, it was also concluded that the acceleration of non-relativistic electrons is only proportional to the acceleration of $>$30 MeV protons when the proton acceleration exceeds a threshold. In fact, it was found that all flares with gamma-ray line emission detectable by RHESSI were accompanied by soft X-ray flares with fluxes greater than 10$^{-5}$ W m$^{-2}$ and that weaker flares without excess of $>$50 keV emission were lacking gamma-ray lines.


RHESSI also made the first imaging of the gamma-ray line emission from solar flares showing that the neutron capture line comes from localized sources similar to the HXR footpoints~\cite{2003ApJ...595L..77H,2006ApJ...644L..93H}. The fact that the imaging of the gamma-ray line emission indicates compact regions and not an extended region suggests that the acceleration (at least during the impulsive phase of the flare) is not due to widespread shocks.  This result together with the correlation between bremsstrahlung and line fluence provides strong evidence that the acceleration of electrons and ions is related to magnetic reconnection.

In order to investigate whether the same acceleration mechanism that is acting on the $>$30 MeV protons is also capable of accelerating protons beyond the $\pi^0$ threshold it is necessary to study the gamma-ray emission at energies greater than 10's of MeV preferably 100's of MeV. Unfortunately the localization accuracy of detectors capable of observing 100's of MeV gamma-ray emission is not as good as the HXR detectors and therefore it is not possible to say whether the $>$100 MeV gamma-ray emission is also located in compact regions close to the HXR footpoints and other methods of analysis are necessary.

\subsection{Beyond the impulsive phase}
\label{sec:beyodimpulsive}
Solar flares with $>$100 MeV emission have also shown to have an intriguing delayed phase lasting for hours after all the other flaring counterpart activity has ceased. In June 1991 the first such detections occurred~\citep{1993A&AS...97..349K, 1996AAS...188.7011R} raising new questions on the origin and acceleration mechanism for this extended emission, leading to the designation of a new class of solar flares: the \emph{Long Duration Gamma-Ray Flares}.  In particular the flare of 1991 June 11 with its 8 hours of gamma-ray emission opened up a new era in gamma-ray solar physics (see right panel of Figure~\ref{fig4} for the time profile of the 1991 June 11 solar flare). At the time of this flare,  two possible scenarios to explain the production and duration of the emission has been envisioned~\cite{1993A&AS...97..349K}: continuous acceleration and diffusion in a turbulent storage loop~\cite{1991ApJ...368..316R} or injections of flare accelerated particles into a \nob{c}oronal loop and subsequent release of particles into the photosphere~\cite{1989ApJ...344..973M,1992ApJ...389..739M}. Given that the electron population of the flare appears to disappear after the first phase of the flare, the first scenario was no longer pursued whereas the second was further considered. Large \nob{c}oronal loops can maintain the particles however complications arise when trying to find a pitch angle scattering value that can allow for the hour-long duration of the emission.

With only a handful of these new \emph{Long Duration Gamma-Ray Flares} observed it was extremely difficult to understand which acceleration mechanism was driving the hour-long emission and where it was occurring. Additional observations of the gamma-ray flaring Sun were needed.

\subsection{Special cases: Behind-the-limb flares}
\label{historybtl}
As already described in section~\ref{sec:intro}, the gamma-ray emission processes require chromospheric densities and the measurements of gamma-ray line emission have been found to be consistent with a compact region located close to the AR from which the counterpart emission originated. However, there is a small subset of special case flares whose AR is located behind the visible solar limb during the flaring activity and yet gamma-ray emission is detected from Earth orbiting satellites. These observations can imply that a spatially extended flare component is present that can subtend a large range of heliolongitudes therefore allowing particles to travel from the acceleration site (close the AR) to the chromosphere on the visible limb. Intense line emission in the 1-8 MeV range and a strong 2.23 MeV neutron capture line from the 1989 September 29 behind-the-limb flare was observed and thus a spatially extended component was considered necessary~\cite{cliv93,1993ApJ...409L..69V}. Another possibility was that the acceleration and emission occur high up in the corona above the AR thus making the emission visible above the limb. This second possibility however does require larger densities than are usually expected for the solar corona if emission such as the neutron capture line are observed. Observations of intense line emission was seen from the 1991 June 1 behind-the-limb flare but no significant neutron capture line and thus a \nob{c}oronal origin for the emission was concluded~\cite{1997ApJ...479..458R,1994ApJ...425L.109B}. The third such flare occurred on 1991 June 30~\citep{1999A&A...342..575V,trottet2003} with emission observed up to almost 100 MeV but no neutron capture line was detected and thus the origin of the emission is still under debate. Clearly with such a limited number of such special case flares it is difficult to confirm the origin and acceleration site of the gamma-ray emission.

\section{\emph{Fermi}-LAT observations of gamma rays from \nob{s}olar flares}
\label{fermiflares}

\begin{figure}
  \includegraphics[width=\textwidth]{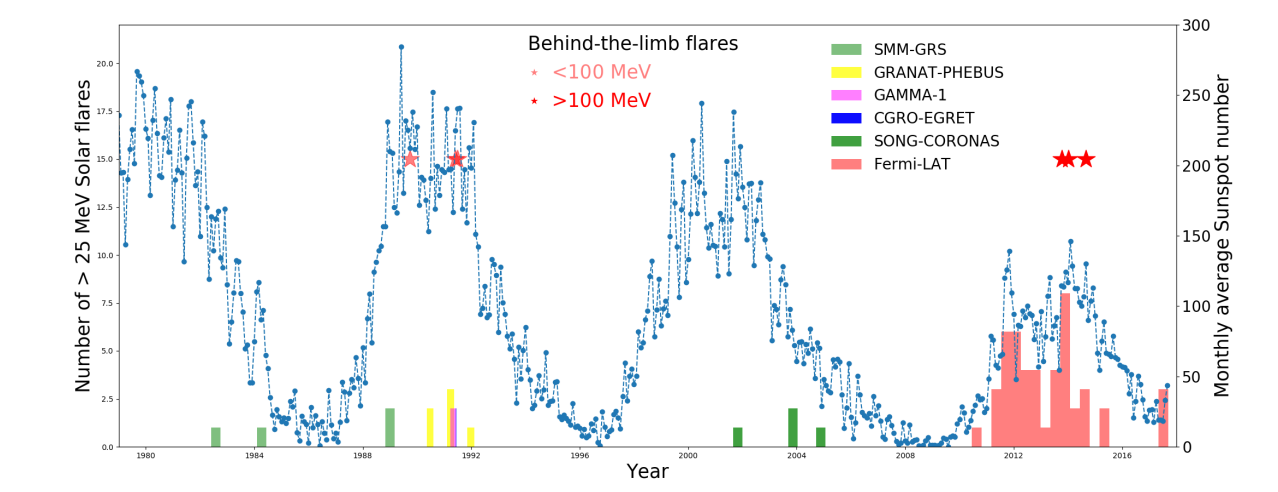}
\caption{Number of flares with emission $>$25 MeV significantly detected since 1980. The histogram is color coded to indicate the telescope that observed the flare, labels are indicated in the upper right-hand legend. The light/dark red stars represent the behind-the-limb solar flares with emission $<$ \nob{or} $>$100 MeV, respectively. The right-hand axis gives the monthly average sunspot number recorder over the years, tracing the solar cycles.}
\label{fig5}
\end{figure}
The LAT observations of the flaring Sun over the 24$^{th}$ solar cycle have revealed an extremely rich and diverse sample of events, spanning from short prompt-impulsive flares~\cite{2012ApJ...745..144A} to the gradual-delayed long-duration phases~\cite{0004-637X-787-1-15} including the longest extended emission ever detected ($\sim$20 hours) from the 2012 March 7 flare~\citep{0004-637X-789-1-20}.  The LAT is an excellent general purpose gamma-ray astrophysics observatory, thanks to its large field of view of 2.4 sr, it monitors the entire sky every two orbits\nob{,} but in doing so it keeps the Sun in the \nob{field of view} only 40\% of the time. Nonetheless, its technology improvements with respect to previous gamma-ray
space based missions, have allowed the LAT to increase the total number of $>$30 MeV detected
solar flares by almost a factor of 10 (see Figure~\ref{fig5} for a comparison of the number of gamma-ray solar flares detected over the past $\sim$40 years). Furthermore, the increase in spatial resolution with respect to EGRET, allowed the localization of the gamma-ray emission centroids on the photosphere for the first time, which is particularly important for the interpretation of the behind-the-limb flares.

Over the first 13 years of the LAT being in orbit it has observed $>$30 MeV emission from 45 flares~\cite{2021ApJS..252...13A}. Of these flares\nob{,} 18 exhibit a prompt component synchronized with the accompanying HXR emission, 37 have emission with a delayed component (where delayed implies not  synchronized with the HXR emission), \nob{eight} have only a prompt component\nob{,} and \nob{three} are behind-the-limb. The flares with a delayed component can be further separated into groups: 21 have delayed emission lasting longer than \nob{two} hours, 16 lasting less than \nob{two} hours and \nob{four} exhibit only delayed emission (no prompt emission was observed). The fact \nob{that} almost half of the entire sample of LAT flares have a significant emission lasting more than \nob{two} hours strongly suggests that such long duration emission is indeed rather common. Another striking aspect of this sample is the fact that roughly half of the flares are associated with moderately bright soft X-ray flares (peak soft X-ray flux $<$10$^{-4}$ W m$^{-2}$, GOES M-class) whereas prior to the LAT all the high-energy gamma-ray solar flares were associated with most intense X-ray flares (peak soft X-ray flux $>$10$^{-4}$ W m$^{-2}$, GOES X-class). Again, suggesting that the process that accelerates ions beyond 300 MeV is fairly common even in modest X-ray flares. The $>$100 MeV peak flux values vary from flare to flare by up to two orders of magnitude, emphasizing the wide variety of the delayed phase of the gamma-ray flares.

The entire LAT sample of solar flares (with the exception of three, see Table 4 in \cite{2021ApJS..252...13A} for further details) are associated with a CME and the majority of these are also fast Halo CMEs. Over the time period covered by the First \emph{Fermi}-LAT \nob{s}olar \nob{f}lare \nob{c}atalog~\cite{2021ApJS..252...13A}, the LASCO CME catalog~\cite{lascocatalog} identified 15841 CMEs with an average linear speed of 342 km\,s$^{-1}$ and maximum speed of 3163 km\,s$^{-1}$. The mean linear speed of the CMEs associated with solar flares with a delayed emission lasting more than two hours is 1766 km\,s$^{-1}$, whereas the flares with only an impulsive phase detected are associated with slower CMEs (mean speed of 656 km\,s$^{-1}$). The connection between fast CMEs and solar flares with hour-long emission was also reported by~\cite{Winter_2018,Share_2018}.

\begin{figure}
  \includegraphics[width=\textwidth]{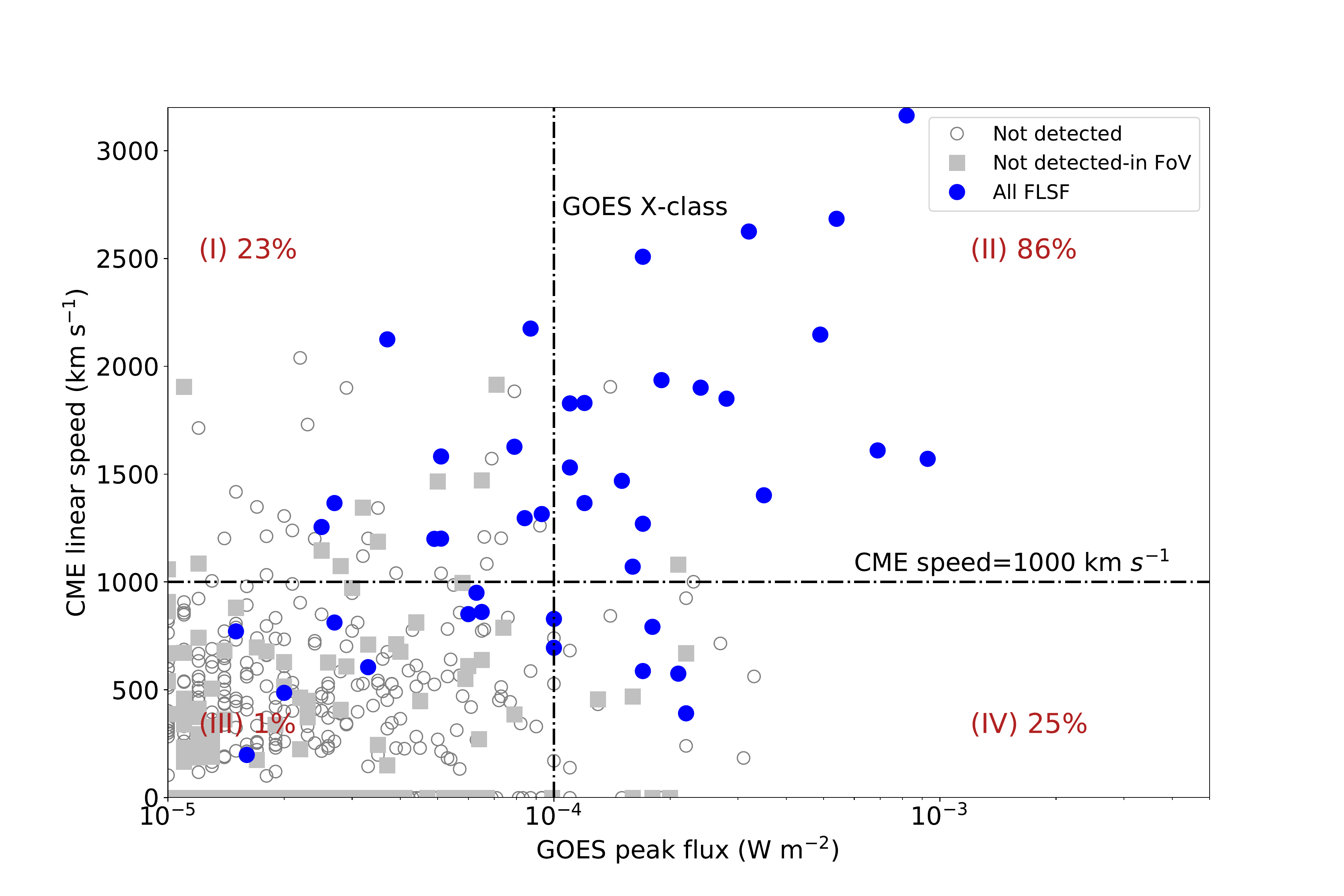}
\caption{CME speed vs soft X-ray peak flux for all flares of Solar cycle 24. The blue markers are the LAT detected flares, the empty circles represent the flare not detected by the LAT and the squares represent the flares not detected by the LAT but that the Sun was in the \nob{field of view} of the LAT at the time of the X-ray flare. In each of the four quadrants (labeled I-IV) the fraction of flares detected by the LAT in that quadrant in annotated in the figure~\cite{2021ApJS..252...13A}. All CME speeds are from the LASCO CME online catalog~\cite{lascocatalog}}
\label{fig6}
\end{figure}

Figure~\ref{fig6} shows a scatter plot of CME speed versus the soft X-ray flux measured by the GOES satellite for all solar flares detected by the LAT (blue marker) and all the M/X class flares non detected by the LAT (gray markers). The dashed vertical and horizontal lines mark four quadrants (I-IV) that indicate the population of flares classified as M/X class and whether they were associated with a CME with linear speed $>$ \nob{or} $<$ 1000 km s$^{-1}$. From this figure it is possible to gather some minimal information on the favorable conditions for gamma-ray emission simply by looking at the fraction of LAT detected flares over the total number of flares that fall within one of the four quadrants. The most favorable condition (86\% of the flares detected) is for flares with SXR flux~$>$10$^{-4}$ W m$^{-2}$ (GOES X-class) and CME with linear speed~$>$1000 km s$^{-1}$, confirming what was at first thought, namely that the gamma-ray emission was associated to the brightest X-ray flares. What is also interesting however, is that quadrants I and IV are equally favorable conditions for $>$30 MeV emission to be detected possibly suggesting that the CME may be playing an equally important role in the acceleration of particles.

\subsection{Localization of the gamma-ray emission}
\label{sec:loc}
\begin{figure}
  \includegraphics[width=.5\textwidth]{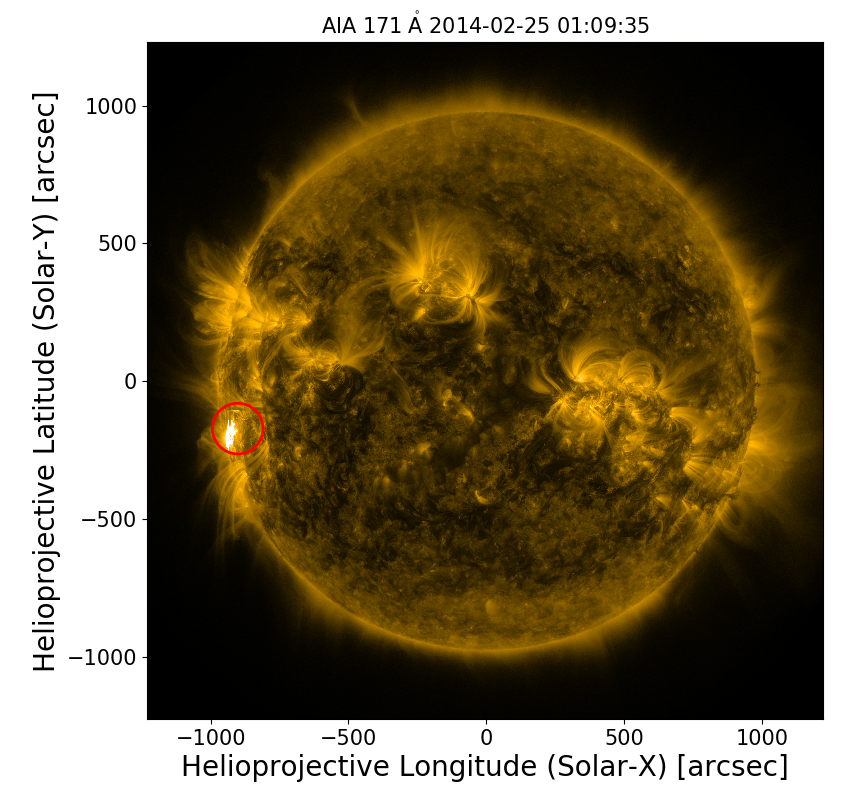}
  \includegraphics[width=.5\textwidth]{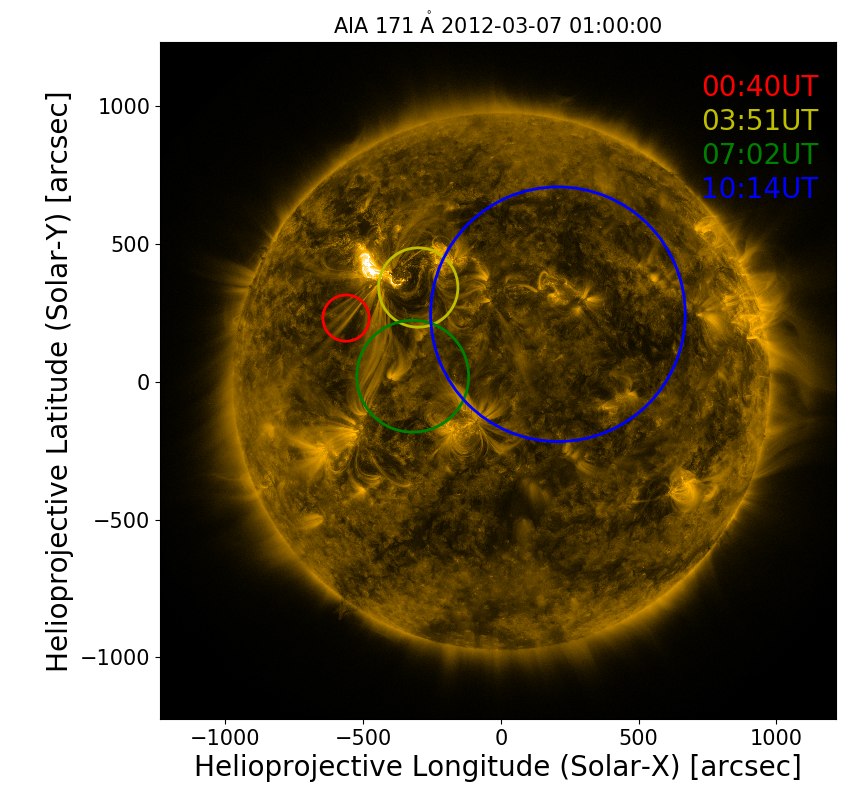}
\caption{Emission centroid for the $>$100 MeV emission observed from the 2014 \nob{February} 25 solar flare(left) and the 2012 \nob{March} 7 flare (right). The flare of 2012 was so intense that the statistics permitted to perform time resolved localization of the first $\approx$10 hours of emission. The localization of the emission from the 2014 flare covers the first 20 minutes of detection. The error radius represents the 68\% containment radius (or 1$\sigma$ statistical localization error) on the position~\cite{2021ApJS..252...13A}.}
\label{fig7}
\end{figure}

The \emph{Fermi}-LAT is the first space telescope capable of determining the centroid of $>$100 MeV emission from solar flares. This is an important additional clue to the overall gamma-ray solar flare puzzle because the location on the solar disk  where the emission is occurring can yield valuable information on where on the photosphere the precipitating ions produce the high-energy gamma rays. Unfortunately, only \nob{eight} out of the 45 flares observed with the LAT where bright enough to yield localization uncertainties $<$365$''$. In Figure~\ref{fig7} there are two examples of solar flares for which the LAT was able to provide localization of the $>$100 MeV emission, the 2014 February 25 (left) and the 2012 March 7 flares (right). The localization of the emission from the 2014 flare covers the first 20 minutes of detection and the emission centroid is centered on the AR which the counterpart flaring emission originated. While the flare of 2012 March 7 was sufficiently intense that the statistics permitted to perform time resolved localization of the first $\approx$10 hours of emission and the localization migrates from the position of the AR (from the early phase of the emission) towards the western limb of the solar disk with time. This result \nob{suggests} that the particle acceleration is occurring at an extended source. \nob{O}n the other hand\nob{,} the result for the impulsive phase of flare of 2014 February 25 seems to support the acceleration at the flare site. Similar results were also found for other impulsive flares observed by the LAT. Further support that the impulsive phase of the emission is occurring at the flare site was found from the 2010 June 12 flare\nob{,}  where it was found that the $>$30 MeV emission lagged the 100 -- 500 keV \nob{bremsstrahlung} emission by 6 $\pm$ 3 seconds indicating that the $>$280 MeV ions were being accelerated on similar timescales as the 100's of keV electrons \cite{2012ApJ...745..144A}. 

\begin{figure}
  \includegraphics[width=\textwidth]{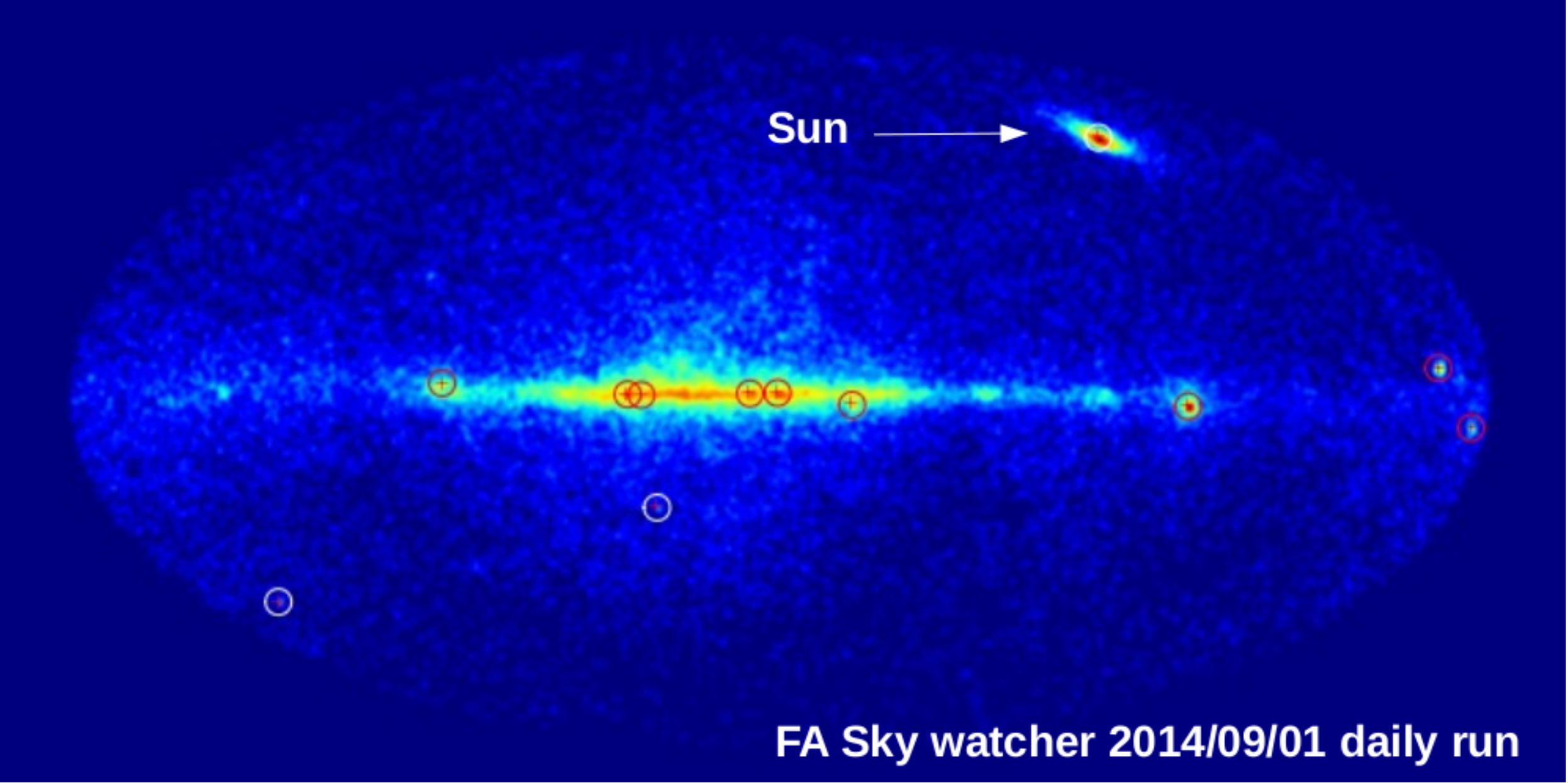}
\caption{The $>$100 MeV sky as observed by \emph{Fermi}-LAT on 2014 September 1$^{st}$. The emission from the solar flare from the AR almost 40$^{\circ}$ behind the visible limb is clearly visible and the intensity of the emission compared to the galactic plane is very evident.}
\label{fig8}
\end{figure}

Of the eight flares for which the LAT was able to provide more accurate localization two  originated from ARs whose position was located behind the visible solar disk. The flare of 2013 October 10 was the first behind-the-limb flare with GeV emission detected (highest energy photon detected was 3.4 GeV), the flare originated from and AR located about 10$^{\circ}$ behind the eastern limb and the LAT observed emission from this flare for roughly 30 minutes\cite{2015ApJ...805L..15P}. The flare of 2014 September 1 originated from an AR almost 40$^{\circ}$ behind the eastern limb and the gamma-ray emission lasted almost two hours (3.5 GeV photons were detected also from this flare)~\cite{2017ApJ...835..219A}. The localization of the emission from these two flares is coincident with the visible disk thus supporting the scenario proposed in~\cite{1993ICRC....3...91C}, namely that a spatially extended component is required in order to explain the observations. Emission in the 1 -- 8 MeV energy range from the 2014 September 1 flare was observed by \emph{Fermi} Gamma-Ray Burst Monitor~\cite{meeg09} but unfortunately no statistically significant line emission was detected. The 2014 September 1 behind-the-limb flare was truly an exceptional flare not only because GeV emission was observed for almost two hours from a flare whose AR was 40$^{\circ}$ behind the visible limb but also because of how intense the emission was. In fact, the Sun was the brightest source in the \emph{Fermi}-LAT sky on the day of the flare with the peak flux reaching more than 1000 times the value of the quiet Sun~\cite{abdo11} level (as can be seen in Figure~\ref{fig8}). 

In order to investigate whether the emission observed from these flare originates from accelerated electrons or ions it is necessary to perform a spectral analysis and test for the various emission models. For the LAT detected flares, the spectral data is fit with three different models. The first two are phenomenological functions that may describe bremsstrahlung emission from relativistic electrons, namely a simple power law (PL) and a power-law with an exponential cut-off (PLEXP). 
When the PLEXP provides a significantly better fit than the PL, a third model that describes pion-decay emission is also considered. The third model uses templates based on a detailed study of gamma rays produced from the  decay of pions originating from accelerated protons with an isotropic pitch angle distribution in a thick-target  model (updated from~\cite{murp87}).

The likelihood ratio test~(TS~\citep{Mattox:96}) is used to estimate the significance of the source (TS$_{\rm PL}$) as well as to estimate whether the addition of the exponential cut-off is statistically significant\footnote{The corresponding difference of maximum likelihoods computed between the two models, $\Delta$TS=TS$_{\rm PLEXP}$-TS$_{\rm PL}$}. The significance in $\sigma$ can be roughly approximated as $\sqrt{\rm TS}$. The pion templates allow to infer the proton index that best describes the data. Once the best proton index is found it is also possible to calculate the estimated total number of protons above a given energy needed to produce the gamma-ray emission observed. This information can be valuable when comparing properties with multiwavelength counterparts of the flare when investigating the origin of the emission.

\begin{figure}
  \includegraphics[width=\textwidth]{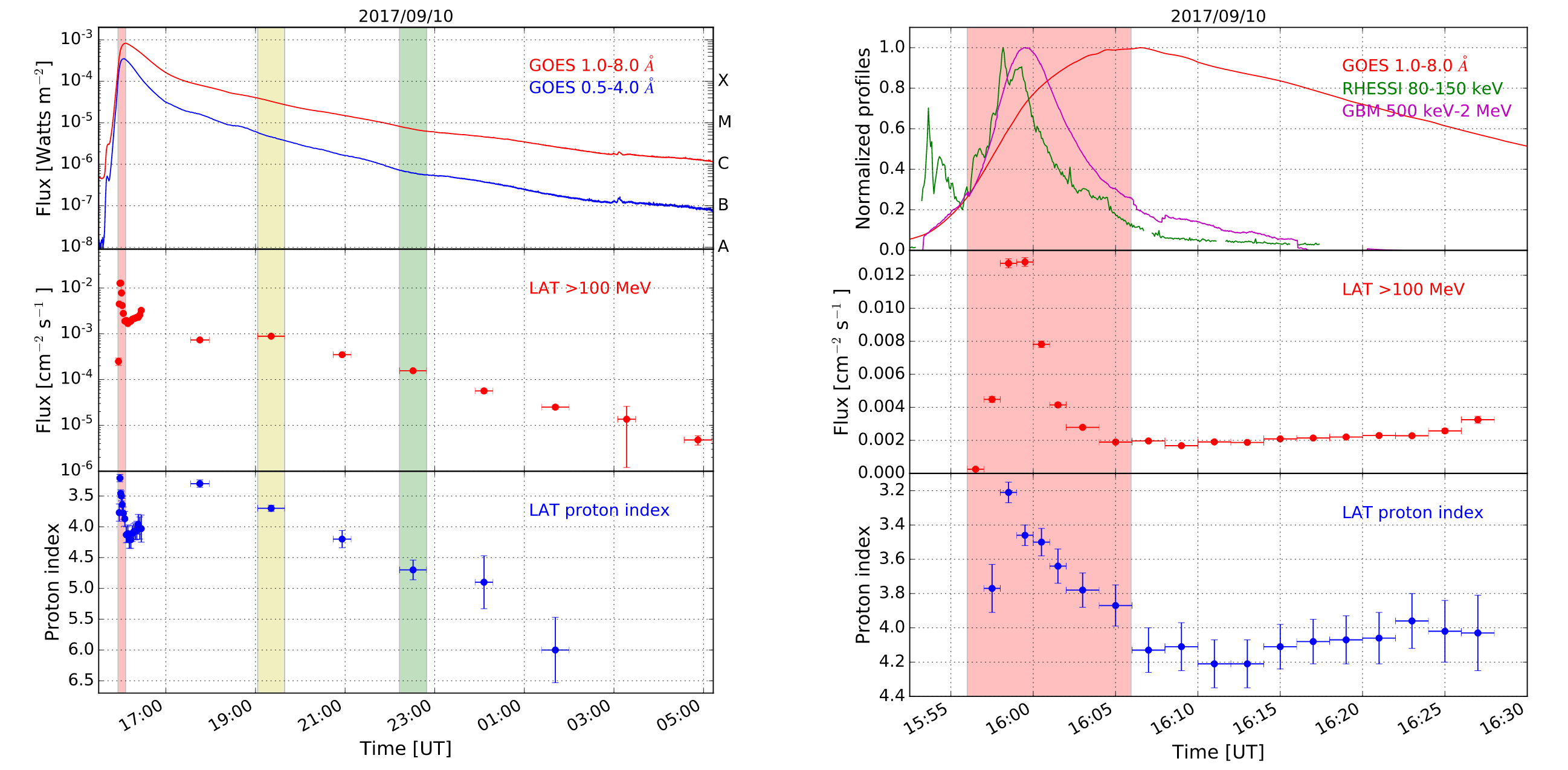}
\caption{Composite light curve for the 2017 September 10 flare with data
from GOES X-rays, \emph{Fermi}-LAT $>$100 MeV flux, and the best proton index
inferred from the LAT gamma-ray data~\cite{2018ApJ...865L...7O}. The full observation time is shown in the left panel and a zoom of the impulsive phase is shown in the right hand panel. The evolution of the proton index shows
three distinct phases, a softening during the prompt-impulsive phase, a plateau and another
softening during the decay phase.}
\label{fig9}
\end{figure}

For all of the flares with sufficient statistics to perform a spectral analysis the data is best described by pion decay and the results for the fits can be found in~\cite{2012ApJ...745..144A}. Given that the LAT is not a solar dedicated mission\nob{,} the coverage of the Sun is not optimal and thus it is not always possible to perform time resolved analyses. For a few cases, where both the prompt and delayed emission occurred when the Sun was in the \nob{field of view} of the LAT and for particular intense flares, it is possible to study not only how the flux varies with time over the various phases\nob{,} but also how the inferred proton index varies with time. One particularly lucky case occurred for the 2017 September 10 flare where the emission lasted for more than 10 hours~\cite{2018ApJ...865L...7O}. The prompt and delayed phase of this flare are shown in Figure~\ref{fig9}. There are several interesting aspects of this flare that can been seen from this figure. For example, the prompt phase of the gamma-ray emission coincides with the X-ray light curve (top and middle panel of the right-hand side of the figure, highlighted in pink) suggesting a common acceleration mechanism during this phase. Following the initial prompt phase, the X-ray light curve decays and while the gamma-ray emission also decays with respect to the prompt, the intensity remains more than 1000 times the quiet Sun level. Furthermore the gamma-ray intensity remains almost constant until the Sun left the field of view ($\approx$20 minutes). When the Sun came back into the field of view, the flux value decreased but continued to display a rise and fall behavior typically seen for the delayed phase. The best proton index also follows a similar behavior as the flux over these three phases of the emission and what is interesting is how the value of the proton index returns to a impulsive phase value ($\sim$3.2) when the delayed phase starts. The data from this flare suggests multiple phases in both the flux and proton index evolution. Similar temporal behavior was also reported by~\cite{rank1991flares} for the extended phase of the first long duration gamma-ray flares observed by COMPTEL in June of 1991. From their detailed analysis of those flares, the authors of~\cite{rank1991flares} concluded that the processes taking place during the extended phase differ from those during the impulsive phase.

 The second Ground Level Enhancement (GLE)\footnote{GLEs are a sudden increases in the cosmic ray intensity recorded by ground based detectors are invariably associated with large solar flares and fast CMEs.} of the 24$^{th}$ solar cycle was also associated with this flare and the properties of this GLE were analyzed in detail by several authors~\cite{Gopalswamy_2018, Kocharov_2020, 2018cosp...42E2298M}. The 24$^{th}$ solar cycle was a GLE poor cycle with only two confirmed events and a handful of sub-GLEs~\cite{2014EP&S...66..104G} as opposed to the 16 observed over the previous solar cycle~\cite{2012SSRv..171...23G} and thus it has not been possible to perform correlation studies between the properties of the delayed gamma-ray emission and those of the GLEs. Given that the CME-driven shock is considered to be the source of large SEPs and GLEs such a study would provide insight on whether the CME shock is also the source of the gamma-ray producing ions. However, several compelling similarities have been found in single sub-GLE events associated with delayed gamma-ray emission, such as the 2014 September 1 flare~\cite{2020SoPh..295...18G}. Whereas the gamma-ray emission associated with the GLE of 2012 May 17 had a peak flux more than two orders of magnitude weaker~\cite{ajello_m_2020_4311157} than the 2017 September 10 flare and the LAT coverage of the $\approx$ 2 hour-long emission was not sufficient to perform any detailed comparison with the GLE.


\subsection{Delayed emission scenarios}

Thanks to the increased sensitivity of the LAT the sample of high-energy gamma-ray solar flares has drastically increased and several additional observational aspects such as localization (both time resolved and time integrated) of the emission  \nob{are now available}. Unfortunately, the sensitivity of the \emph{Fermi}-GBM to nuclear lines is not as optimal when compared to detectors such as SMM and thus a complete picture of how and where the ion population is being accelerated during solar flares is still missing. From the $>$100 MeV observations provided by the LAT two main competing scenarios have surfaced in the solar physics community to explain the hour-long delayed emission: acceleration at the CME-driven shock with back precipitation\nob{,} or acceleration via second-order Fermi mechanism and trapping locally within extended coronal loops.

Several observational clues have been reported supporting the CME shock scenario, for example the duration of the $>$100 MeV gamma-ray solar flares show\nob{s} a linear relationship with the duration of the associated type II radio bursts\nob{,} and that the number of protons observed at Earth as Solar Energetic Particles (SEP) and the number of protons needed to produce the gamma rays also show a linear correlation~\cite{Gopalswamy_2021,Gopalswamy_2019}. These two results strongly \nob{suggest} a common acceleration mechanism between these two phenomena\nob{,} and that the protons responsible for the gamma-ray emission and the SEPs are drawn from the same population. Additional support from data-driven global magnetohydrodynamic simulations of the 2014 September 1 behind-the-limb solar flare was also reported in~\cite{Jin2018}. In their work, they found that magnetic connectivity was established between the CME-driven shock and the centroid of the gamma-ray emission\nob{,} and that the rate of particle acceleration by the shock closely correlated with the gamma-ray flux.  Further support for the CME-shock scenario has been reported by~\cite{2018ApJ...868L..19G,2020SoPh..295...18G,Kouloumvakos_2020,Plotnikov_2017,Share_2018}.

On the other hand, the authors of~\cite{Share_2018} and \cite{DeNolfo2019} did not find a correlation between the  number of protons observed at Earth as SEP and the number of protons needed to produce the gamma rays. There is a large range of systematic uncertainties tied to estimating the total number of protons both from SEPs and from gamma-ray emission\nob{,} and further investigation is needed to resolve the contrasting results between \cite{Gopalswamy_2021} and \cite{DeNolfo2019,Share_2018}. The authors of~\cite{DeNolfo2019} suggest an alternative scenario where the protons that have been accelerated are trapped locally and proceed to diffuse to the denser photosphere to radiate~\cite{1991ApJ...368..316R}. Therefore\nob{,} the two populations of protons are no longer intimately connected. Observations of large quasi-static loops in gyrosynchrotron emission by accelerated electrons from the Nancay Radioheliograph of the 2014 September 1 flare were reported by~\cite{Grechnev_2018}. In their work they compared time profiles in several wavelengths and found them to be consistent with prolonged confinement of particles injected within a magnetic trap. Large extended loops were also observed in microwave observations from EOVSA of the 2017 September 10 flare~\cite{Gary_2018} providing support for the \nob{c}oronal trap scenario. However it is not clear if these two observations one on emission by protons and one on emission by electrons are related. Additional joint observations of HXR and microwave emission is needed to support this scenario and overcome the complications discussed in~\cite{2016ApJ...830...28P}.

\section{Conclusions}
The increase in the sample of gamma-ray solar flares provided by LAT has helped in expanding the understanding of how and where particles are being accelerated during these explosive phenomena. However\nob{,} several questions still remain unanswered and the \emph{smoking gun} evidence in support of a delayed emission scenario has yet to be observed. As always, further observations with better precision \nob{are} needed to help in answering the questions still \nob{puzzling} the solar physics community. New observatories such as (and not limited to) EOVSA\footnote{\url{http://www.ovsa.njit.edu/}}, Parker Solar Probe\cite{2016SSRv..204....7F} and Solar Orbiter~\cite{2021A&A...646A.121G} will certainly provide valuable additional information that will help to further our understanding of the flaring Sun.

\bibliographystyle{JHEP}
\bibliography{pescerollins_solarflares}


%
%
%

\end{document}